\documentclass[a4paper,11pt]{article}
\pdfoutput=1
\usepackage{amsmath}
\usepackage[colorlinks=true]{hyperref} 
\usepackage{graphicx,rotating,hyperref,slashed,amsmath,xcolor,amssymb,amsfonts,colortbl,cite} 
\usepackage{microtype}
\makeatletter 
\hypersetup{colorlinks,bookmarksopen,bookmarksnumbered,  
linkcolor=blus,pdfstartview=FitH,urlcolor=rossos,citecolor=verde}
\allowdisplaybreaks		 

\hypersetup{%
    ,urlcolor=black
    ,citecolor=black
    ,linkcolor=black
    }
\usepackage{mciteplus}

\AtBeginDocument{
  \hypersetup{
    urlcolor=blue,
    citecolor=blue,
    linkcolor=black,
  }%
}

\usepackage{colortbl}
\definecolor{green3}{cmyk}{0.8, 0., 0.7., 0.}
\definecolor{green2}{cmyk}{0, 1, 0.5, 0}
\definecolor{lightgreen}{cmyk}{0.2, 0, 0.2, 0.2}
\definecolor{lightgray}{cmyk}{0.1,0.2,0,0.1}
\definecolor{lightgray2}{cmyk}{0.4,0.4,0,0.8}
\definecolor{black}{cmyk}{1.0,1.0,1.0,1.0}

\allowdisplaybreaks[1]

\usepackage{colortbl}
\definecolor{lightgreen}{cmyk}{0.2, 0, 0.2, 0.2}
\definecolor{lightgray}{cmyk}{0.1,0.2,0,0.1}
\definecolor{lightgray2}{cmyk}{0.1,0.1,0,0.1}

\makeatletter
\newlength{\apb@width}
\newcommand{\autoparbox}[2][c]{\settowidth{\apb@width}{#2}\parbox[#1]{\apb@width}{#2}}

\makeatother

\numberwithin{equation}{section}

\def\be{\begin{equation}}
\def\ee{\end{equation}}

\def\beq{\begin{equation}}
\def\eeq{\end{equation}}

\def\bea{\begin{eqnarray}}
\def\eea{\end{eqnarray}}

\def\0{{\boldsymbol 0}}

\usepackage{setspace} 

\begin{document}

\begin{titlepage}

\setcounter{page}{1} \baselineskip=15.5pt \thispagestyle{empty}

\bigskip\

\vspace{1cm}
\begin{center}

{
{\fontsize{19}{28}\selectfont  \sffamily \bfseries {
{Stochastic approach to
 gravitational waves  \\ \vskip0.2cm from inflation}
}}
}
\\
\hskip1cm

\end{center}

\vspace{0.2cm}

\begin{center}
{\fontsize{13}{30}\selectfont  Gianmassimo Tasinato } 
\end{center}

\begin{center}

\vskip 8pt
\textsl{ Physics Department, Swansea University, SA28PP, United Kingdom }\\
\vskip 7pt

\end{center}

\vspace{1.2cm}
 \vspace{0.3cm}
\noindent {\sffamily \bfseries Abstract} \\[0.1cm]    
We propose a coarse-graining procedure for describing the superhorizon
dynamics of inflationary tensor modes. Our aim is to formulate a   stochastic
description for the statistics of spin-2 modes  which seed the background of gravitational
waves from inflation. 
   Using basic principles of
quantum mechanics, we determine a probability density for 
coarse-grained tensor fields, which satisfies a stochastic Fokker-Planck
equation at superhorizon scales. The corresponding noise and drift are computable, and depend on
the cosmological system under consideration. Our general formulas are
applied to a variety of cosmological scenarios, also considering   cases seldom
considered in the context of stochastic inflation, and which are important for their observational consequences. We start obtaining
the expected expressions for noise and drift in pure de Sitter and
power-law inflation, also  including a 
discussion of effects of 
non-attractor phases.    We  then  
 apply our  methods to describe scenarios with a transition from inflation to standard
 cosmological 
  eras of radiation and matter domination.
    We show how  the
interference  between
modes flowing through the cosmological horizon,  and modes spontaneously
produced  at superhorizon scales, can affect the stochastic evolution
of coarse-grained tensor quantities.
 In  appropriate limits,  we find that the corresponding
spectrum of tensor modes at horizon crossing matches with the results of quantum field theory calculations, but we also highlight where differences can arise.

\vspace{0.6cm}
 \end{titlepage}

\section{Introduction}

Cosmological inflation is the most successful  mechanism at our disposal for   
 generating the initial conditions for our universe \cite{Guth:1980zm,Albrecht:1982wi,Linde:1981mu,Mukhanov:1981xt,Guth:1982ec,Hawking:1982cz,Starobinsky:1982ee}. 
 During cosmological inflation,  space-time
 fluctuations produced by quantum effects at microscopic distances 
are stretched to superhorizon
scales, where they freeze.
Subsequently,  after inflation ends, such large-scale fluctuations reenter the horizon, 
 they
become
dynamical,  and seed
the  evolving cosmic structures we observe today in the sky.

\smallskip

This picture of early universe cosmology is  appealing and physically well motivated. Nevertheless,  potentially large infrared effects require 
specific care  when  applying pertubative  quantum field theory techniques  to cosmology. For example, it is well known \cite{Ford:1984hs,Antoniadis:1985pj,Tsamis:1994ca} that perturbative computations of correlation functions of light quantum fields in de Sitter space can be 
affected by  infrared  contributions,   making  subtle  a proper physical interpretation of the calculations: see e.g.     \cite{Urakawa:2009my,Giddings:2010nc,Byrnes:2010yc,Burgess:2010dd,Gerstenlauer:2011ti,Tanaka:2017nff}. A promising proposal to deal with these issues is the stochastic approach to cosmological inflation first proposed by Starobinsky \cite{Starobinsky:1986fx}, which   
 provides a consistent framework for resumming large infrared effects in de Sitter space: see e.g. \cite{Nambu:1987ef,Kandrup:1988sc,Nambu:1989uf,Mollerach:1990zf,Linde:1993xx,Starobinsky:1994bd,Wands:2000dp,Finelli:2008zg,Garbrecht:2013coa,Burgess:2014eoa,Vennin:2015hra,Burgess:2015ajz,Hollowood:2017bil,Pattison:2019hef}.  The starting point of  stochastic inflation is  the observation that  
 after crossing the cosmological
 horizon, quantum fluctuations classicalize \cite{Brandenberger:1990bx,Albrecht:1992kf,Polarski:1995jg,Calzetta:1995ys,Lesgourgues:1996jc,Kiefer:1998qe,Lombardo:2005iz,Burgess:2006jn,Kiefer:2008ku}, and their description is more conveniently formulated
 in terms of a classical, stochastic Fokker-Planck evolution equation.  
In this perspective,    long wavelength modes at superhorizon scales  receive    
  impulses  from  small-scale fluctuations as the latter 
  cross
 the   horizon, a  process  that intuitively corresponds to  a cosmological version of Brownian motion. The resulting stochastic cosmological equations can be handled consistently,  and provide information on the global dynamics of the system at the largest cosmological scales, which is  difficult to gain  otherwise.

\smallskip

In this work  we study  the stochastic distribution of cosmological  fluctuations 
at superhorizon scales, 
 focusing on the dynamics of primordial tensor modes predicted by inflation \cite{Grishchuk:1974ny,Starobinsky:1979ty,Starobinsky:1980te,Grishchuk:1990bj}. Spin-2 inflationary tensor modes
 are light fields in quasi-de Sitter space:  their superhorizon distribution is likely  to be amenable of    a classical    description as in
the stochastic approach to scalar fluctuations during cosmological inflation. 
%
 Specifically, we aim to address  two questions: 
\begin{itemize}
\item {\bf Question 1:} Is there a  way to define coarse-grained  tensor modes at superhorizon scales, and study their corresponding dynamics using a stochastic approach?

 A
 reliable stochastic formalism applied to inflationary spin-2 fields  would  allow us to discuss the  dynamics of superhorizon tensor modes using  statistical methods, 
 without having to make specific assumptions on the behavior of the  individual modes after they cross the horizon during inflation.

%

\item {\bf Question 2:}  Is there a consistent  stochastic description for superhorizon inflationary tensor modes {\it after} the end of inflation?

This issue has important implications for cosmology, since 
after inflation ends tensor modes reenter the horizon forming the stochastic background
of primordial gravitational waves  currently  searched by dedicated experiments. Their properties depend on the amplitude and properties of the spectrum at horizon crossing, which depend on the stochastic distribution of tensor modes at the largest, superhorizon scales. 
\end{itemize}
The answer to {\bf Question 1} requires a 
definition of tensor `zero-modes' which  can be subtle  since superhorizon spin-2 fields  do not preserve the isotropy of the  underlying  Friedmann-Robertson-Walker (FRW) space-time. 
Starting from  basic principles of quantum mechanics, in section \ref{sec_form}
 we propose     a coarse-grained  description
of primordial tensor modes, based on the method of the functional
 Schr\"odinger picture \cite{Guth:1985ya} used
in \cite{Burgess:2014eoa,Burgess:2015ajz} for  a stochastic analysis of the scalar sector
of fluctuations. The coarse-grained tensor  quantity we
define is representative of the dynamics of long wavelength tensor
 modes once they leave the cosmological horizon. Its definition does not interfere with the symmetries
 of the background space-time. In fact,
     we focus on a free 
theory described by a quadratic tensor action, with the specific purpose of understanding
how the properties of the coarse-grained quantities depend on the curved cosmological
space-time where they are embedded.
  We obtain
   a probability density for the coarse-grained   
superhorizon tensor modes, and we derive its corresponding   classical Fokker-Planck evolution equation.  It is built
in terms of  
 noise and drift, which  are explicitly calculable  from  combinations
of mode functions  evaluated at superhorizon scales. The definitions of noise and drift are free
from large infrared effects. The noise
is induced by a flow of modes as they cross the cosmological horizon
from small to large scales (or vice versa) -- as in the aforementioned  cosmological
analog of Brownian motion. But it    can also be affected by 
phenomena  occurring   beyond  horizon crossing scales, as for example    interference among the flow of modes  with    
particles produced at superhorizon scales by sizable space-time gradients.

The evolution of  the coarse-grained 
probability density is Markovian, up to   
contributions associated with modes that rapidly decay at superhorizon
scales. Such   effects make the structure of evolution equations dependent on initial conditions,  but they are
negligible in scenarios where cosmological evolution is an attractor.  However, they
can provide a sizable contribution to the drift term in scenarios that
include phases of non-attractor evolution, and our general formulas can be applied
to those set ups as well.  

We also derive formulas for the spectrum of tensor fluctuations evaluated
at horizon crossing, which is useful for comparing with results from QFT computations. Moreover, since we are dealing with 
 coarse-grained quantities, we can define a Gibbs entropy for the system at 
 superhorizon scales. We find that it increases with the universe expansion, and
 we quantitatively  characterize its growth.

\medskip

As far as we are aware,   we are the first in attempting to answer  {\bf Question 2} 
  %
%
  %
in the
context of a stochastic description of superhorizon tensor modes. 
 %
 %
 %
 We start
 in section \ref{sec_puredS} with the case of inflation:
 we recover
the   expected results for the stochastic  distribution of coarse-grained tensor fields during
 de Sitter and power-law cosmological expansion. We also consider the case for  
 non-attractor cosmological evolution, showing explicitly how it  affects
 the drift contributing  to the Fokker-Planck equation.  In section \ref{sec_power} we  apply our stochastic formulas
  to the case of radiation and   matter dominated eras 
 occurring after
 inflation ends.  These stochastic equations describe the coarse-grained evolution of superhorizon tensor 
 modes that  eventually reenter the horizon as cosmic evolution proceeds. 
 The computation of the stochastic noise  makes manifest   
  interference effects among 
  the flow of modes reentering the horizon after inflation ends,   and the superhorizon      
modes semiclassically  produced at large scales by  large space-time gradients, see e.g.
   \cite{Grishchuk:1974ny,Starobinsky:1979ty,Abbott:1985cu,Ford:1986sy}.
 The formula for the noise depends on the number of e-folds of cosmic expansion, and it rapidly
 approaches   a constant value after a few e-folds. Also, 
we prove that our final
results do not depend on the choice of  infrared cutoff, the latter providing contributions   that are exponentially suppressed by the number of e-folds of expansion. Our stochastic formalism can then be used to compute the spectrum of tensor fluctuations at horizon exit, that in appropriate limits
coincide with   the results of  QFT calculations. 

\medskip

Our work aims 
 to put in a firmer footing the intuitive idea that the stochastic distribution of tensor
fields at superhorizon scales is due to the flow of modes between subhorizon and superhorizon scales.  
 A  general lesson of our approach is that   a classical, stochastic approach to primordial tensor fluctuations from inflation is feasible and provides new physical insight in cosmological  situations not usually considered in a stochastic  context. Our
 results are consistent  expectations from a traditional QFT approach to cosmological fluctuations from inflation. It can be used for better clarifying the classical dynamics of tensor modes at large superhorizon scales, and for dealing with large infrared effects from long wavelength modes. We summarize and further discuss physical implications of our results
 in section  \ref{sec_conc}, which is  followed by a technical appendix \ref{appcogr}. Throughout this work we  set  $\hbar\,=\,c\,=\,1$.

\section{  A Fokker-Planck equation for tensor modes from inflation}
\label{sec_form}

After crossing the cosmological horizon, single-field  inflationary scalar and tensor fluctuations 
become time independent, and their spatial configurations 
 can be described in terms of    classical, but stochastically distributed 
 superhorizon modes.  
 
 \smallskip
 
 Our aim in this section is to discuss a  systematic method for obtaining
the classical  evolution equation describing  stochastic, coarse-grained 
superhorizon modes, starting from  basic principles of quantum mechanics. We discuss  free theories in arbitrary  cosmological backgrounds equipped with  a cosmological horizon, with the specific aim of extracting
 the effects of curved space on the derivation of the stochastic equation.
  We concentrate
 on tensor fluctuations, being
the stochastic approach for scalar fluctuations already well developed (including the effects of self-interactions).

\smallskip

  For 
 determining 
  the desired stochastic equation, we 
  make use of the approach of \cite{Burgess:2014eoa,Burgess:2015ajz} based on
 the Schr\"odinger functional picture, first applied to inflationary cosmology in \cite{Guth:1985ya}.
  (See instead \cite{Morikawa:1989xz,Hu:1994dka,Matarrese:2003ye} for derivations of inflationary stochastic equations using  a Schwinger-Keldish approach.)
  We start 
 in subsection \ref{sub_system} setting the stage for the system we consider, and 
 reviewing how the  Schr\"odinger  formalism leads to an evolution equation for probability densities associated
 with quantum Fourier modes of inflationary fluctuations. 
 In subsection \ref{sub_coarse} we define the coarse-grained superhorizon quantities we consider, and we  derive the
 classical stochastic evolution 
 equation for the distribution  of the coarse-grained quantities. The result is a  Fokker-Planck
 evolution equation, for which  we provide the  expressions for  noise and  drift.
 In subsection \ref{sub_spectrum} we discuss how to use these results for computing the spectrum of tensor fluctuations evaluated
 at horizon crossing, as well as the Gibbs  entropy associated with superhorizon coarse-grained tensor modes. 
  

\subsection{  The system we consider}
\label{sub_system}

We consider a space-time described by a  conformally flat FRW metric perturbed by  spin-2  tensor perturbations: 
\be \label{metr1a}
d s^2\,=\,a^2(\tau)\,\left[- d \tau^2+\left[\delta_{ij}+h_{ij}(\tau,\,\vec x) \right]\,d x^i d x^j \right]\,,
\ee
where $a(\tau)$ is the scale factor, while $h_{ij}$ denotes the linearized, transverse-traceless tensor fluctuation,  
 gauge invariant
at first order in perturbations. 
We do  not need to specify  the explicit time dependence  of the  scale factor for developing
our arguments, which can then be applied to a variety of cosmological setups (see section \ref{sec_appl}).  
 The
effective quadratic action controlling the  tensor modes in eq \eqref{metr1a} is
\be \label{quadr1}
S_h^{(2)}\,=\,\frac{M_{\rm Pl}^2}{8}\,\int d \tau\, d^3 x\,a^2(\tau) \left[h_{ij}'^2 - (\vec \nabla  h_{ij})^2\right]\,,
\ee
where  prime denotes derivative along conformal time. 

We express tensor fluctuations in Fourier space, defined within a box of comoving size $L$ (in due time we will consider the  limit of infinitely  large box size):
\be\label{FouEx1}
h_{ij}(\tau, \vec x)\,=\,\frac{2}{M_{\rm Pl}\,L^{3}}\,\sum_\lambda\,\sum_{\vec k}\,h^{(\lambda)}_k(\tau)\,{\bf e}_{ij}^{(\lambda)} (\hat k)
\,e^{i \vec k \vec x}\,,
\ee
with $\vec k\,=\,k\,\hat k$ the tensor 3-momentum, and $\lambda$ its  polarization. ${\bf e}_{ij}^{(\pm)} (\hat k)$ are  (real) helicity
tensors normalized as (we sum over repeated spatial indexes)
\be
{\bf e}_{ij}^{(\lambda)} (\hat k)\,{\bf e}_{ij}^{(\lambda')} (\hat k)\,=\,2\,\delta^{\lambda \lambda'}\,.
\ee
To ensure that   $h_{ij}(\tau, \vec x)$ is real, we demand $\left(h^{(\lambda)}_k(\tau)\right)^*\,=\,h^{(\lambda)}_{-k}(\tau)$,
and in writing eq \eqref{FouEx1}  we sum over positive as well as  negative values of $k$. Plugging 
eq \eqref{FouEx1} in \eqref{quadr1}, we find the quadratic action for mode of momentum $k$:
\be
S_k\,=\,\sum_\lambda \int d \tau\,a^2(\tau) \left[h'^{(\lambda)}_{k}\,h'^{(\lambda)}_{-k}  - k^2  h^{(\lambda)}_{k} h^{(\lambda)}_{-k} \right]\,.
\ee 
The associated Lagrangian density ${\cal L}_k$ is the argument of the previous integral, and allows
us to define the momentum
\be
\pi_k^{(\lambda)}\,\equiv\,\frac{\delta {\cal L}_k}{\delta\,h'^{(\lambda)}_{k} } \,=\,a^2(\tau)\,h'^{(\lambda)}_{-k}\,.
\ee
This information can  be used to obtain the 
Hamiltonian density
\be
{\cal H}^{(\lambda)}_k\,=\,\frac{1}{a^2(\tau)}\,\pi_k^{(\lambda)} \pi^{(\lambda)}_{-k}+a^2(\tau)\,k^2\,h^{(\lambda)}_k\,h^{(\lambda)}_{-k}
\,,
\ee
which is 
a basic ingredient for our next discussion.

\subsubsection*{The functional  Schr\"odinger picture }

We make use of the functional  Schr\"odinger picture 
to derive the evolution equation for probability densities for the system under consideration. 
 We apply the approach previously developed in  \cite{Burgess:2014eoa} to the case of spin-2 tensor modes.  In this subsection we make use
of basic rules of quantum mechanics; in the next subsection \ref{sub_coarse} we show how  an appropriate
coarse-grained procedure  leads to  a classical, stochastic evolution equation for the superhorizon quantities
we are interested in. 

\smallskip

In the functional  Schr\"odinger formalism,  the quantities $h^{(\lambda)}_k$ and $\pi_k^{(\lambda)}$
  are promoted to operators  $\hat h^{(\lambda)}_k$ and $\hat \pi_k^{(\lambda)}$.
An abstract quantum mechanical state in Fourier space
is realized by $\Psi^{(\lambda)}_k \left(h^{(\lambda)}_k, \tau \right)$  which is a
wave functional of the $c$-number quantity  $h^{(\lambda)}_k$,  and it is evaluated 
at a  time $\tau$.  
The action of the operator $\hat h^{(\lambda)}_k$ on the quantum
state is realized by multiplying  $\Psi^{(\lambda)}_k $ by $h^{(\lambda)}_k$, 
while the action of the canonical momentum
$ \hat \pi_k^{(\lambda)} $ is realized by functional differentiation:
\bea
\hat h^{(\lambda)}_k\,| \Psi^{(\lambda)}_k \rangle&\to&h^{(\lambda)}_k\,\,\Psi^{(\lambda)}_k\,,
\\
\hat \pi_k^{(\lambda)} \,\,| \Psi^{(\lambda)}_k \rangle&\to&\frac{1}{i}
\frac{\partial\,\Psi^{(\lambda)}_k}{\partial h^{(\lambda)}_k}\,.
\eea
The Schr\"odinger formalism dictates that for each mode $k$ and polarization $\lambda$
the evolution of the quantum state is controlled by the Schr\"odinger equation
\be\label{schroEQ}
i \,\frac{\partial\,\Psi^{(\lambda)}_k (\tau)}{\partial\,\tau}\,=\,{\cal H}^{(\lambda)}_k\,\Psi^{(\lambda)}_k(\tau)\,,
\ee
with Hamiltonian
\be
{\cal H}^{(\lambda)} _k\,=\,-\frac{1}{a^2(\tau)}\,\frac{\delta^2}{ \delta  h^{(\lambda)}_k\,\delta  h^{(\lambda)}_{-k}}
+a^2(\tau)\,k^2\,  h^{(\lambda)}_k\,  h^{(\lambda)}_{-k}\,.
\ee
Ours is a free theory, and
we can use a Gaussian Ansatz for parametrizing the  wave function. We assume no parity violation, hence the 
explicitly time-dependent functions appearing in our Ansatz are assumed not to 
 depend on the polarization index $\lambda$:
\be \label{quadANps}
\Psi^{(\lambda)}_k\left[ h^{(\lambda)}_k,\,\tau\right]\,=\,{ \Omega}_k(\tau)\,\exp{\left\{ -a^2(\tau) \left[ \alpha_k (\tau) 
 h^{(\lambda)}_k\,  h^{(\lambda)}_{-k}
-\beta_0(\tau) \,\delta_{k0}\, h^{(\lambda)}_k
\right] \right\}}\,.
\ee
The  zero-mode contribution proportional to   $\beta_0$  is not forbidden hence we need to include it -- as we will 
see it is relevant when discussing the effects of the zero mode of infinitely large
wavelength. 

Plugging Ansatz \eqref{quadANps} in \eqref{schroEQ}, 
the
system of equations to solve is (all quantities a part from $k$ depend on time $\tau$)
\bea
0&=&{\Omega}_k'+i\,\alpha_k\,{\Omega}_k\,,
\\
\label{eqak}
0&=&\alpha_k'+i \,\alpha_k^2+\frac{2 a'}{a}\,\alpha_k-i k^2
\,,
\\
0&=&\beta_0'+i \,\alpha_0\,\beta_0+\frac{2 a'}{a}\,\beta_0
\,.
\eea
Combining the last two equations, we find the relation $\beta_0(\tau)\,=\,{\cal C}_\beta\,\alpha_0(\tau)$, with ${\cal C}_\beta$
arbitrary constant (that will not enter in our final results). To deal with eq \eqref{eqak}, it is convenient to define  
\cite{Guth:1985ya} 
\be\label{defak}
\alpha_k(\tau)\,=\,\frac{1}{i}\,\partial_\tau\,\ln\left[ \frac{\gamma^\star_k(\tau)}{a(\tau)}\right]\,.\hskip1cm
\ee
Plugging in eq \eqref{eqak} we get a second order, linear equation for $\gamma_k$ 
\be\label{eqphk1}
\gamma_k''+\left( k^2-\frac{a''}{a} \right)\,\gamma_k\,=\,0\,.
\ee
Since 
 the definition \eqref{defak} involves  derivatives of a logarithm, we can choose the preferred normalization
 for the mode  $\gamma_k$. We impose  the  Wronskian condition
\be \label{wroco1}
\gamma_k'\,\gamma_k^\star
-\gamma_k'^\star\,\gamma_k\,=\,i\,.
\ee
Given these conditions, the following relations hold
\bea
\alpha_k+\alpha_k^\star&=&-\frac{1}{|\gamma_k|^2}\,,
\\
\alpha_k-\alpha_k^\star&=&
\frac{1}{i}\,\partial_\tau\,
\ln
\left[ 
\frac{|\gamma_k|^2}{a^2(\tau)}
\right]\,.
\eea
For any $k\neq0$, we can impose the Bunch-Davies initial conditions
 at early times $\tau\to-\infty$, since at  very small scales the effect of  space-time curvature 
 can be neglected.
 As shown in \cite{Guth:1985ya}, this is equivalent to
 ensure that the wave function at early times is the one of a harmonic oscillator. These conditions
 completely  fix the solution for each mode $k\neq0$.
 
 \smallskip

We need special care in dealing with the zero mode. 
 In this case 
the  Bunch-Davies  condition  does not apply, since $k=0$
can never acquire a small-scale limit for any given time $\tau$. 
The Fourier mode $k=0$  is a linear combination  of the  two independent solutions 
\be
 \gamma_0(\tau)\,\propto\,
a(\tau)\,\,\,\,,  \,\,\,\,\gamma_0(\tau)\,\propto\,a(\tau)  {\cal I}(\tau) \,,
\ee
which 
 solve eq \eqref{eqphk1}. The quantity  ${\cal I}(\tau)$  is defined as
 \bea\label{intItau}
 {\cal I}(\tau)&=&\int_{\tau_\star}^{\tau}\,
 \frac{d \tilde \tau}{a^2(\tilde \tau)}\,,
 \eea
 with $\tau_\star$ an arbitrary fiducial time.
 We find  convenient to express the zero mode  as
 \bea
 \label{newezm1}
\frac{ \gamma_0(\tau)}{ a(\tau) }&=&
  \sqrt{\frac{1}{2\,\mu\, \sin{(\Delta \theta)}}}+\sqrt{\frac{\mu\,e^{2\,i\, \Delta \theta}}{2 \sin{(\Delta \theta)}}}\,{\cal I}(\tau) \,,
 \eea
  with $\Delta \theta$, and $\mu$ two arbitrary real quantities.  
  Their values
 can be associated with the initial conditions on the zero mode at fiducial time $\tau_*$.     We will  study in what comes next how the dependence on initial conditions affects the structure of the evolution
 equations for the quantities we are interested in. 
 
 The 
  expression \eqref{newezm1} automatically satisfies  the Wronskian condition. An overall phase
    can be included in the zero-mode solution, but it has no physical consequences. 
  The
 solution for $\alpha_0$, as defined in \eqref{defak}, reads 
\be\label{sola0}
\alpha_0(\tau)\,=\,\frac{
\mu\,e^{-i \left(\Delta \theta+ {\pi}/{2} \right)}}{a^2(\tau)}\,\frac{1}{1+
\mu\,e^{-i \Delta \theta}\,{\cal I}(\tau)}\,.
\ee
It depends  on  $\mu$ and $\Delta \theta$, determined by the initial conditions
at time $\tau_*$. But notice that the value of $\alpha_0(\tau)$ depends 
 also on the integral ${\cal I}(\tau)$ which depends
on the entire cosmological history from the fiducial initial time $\tau_\star$ to $\tau$. 

  \smallskip
  Once we have control on the quantities entering in the wave functional,
  we  define a probability density associated with the quantum state of momentum  $k$. As usual in quantum mechanics, this quantity  is  
  proportional to the square of the wave functional \footnote{We are focusing on the diagonal elements of the density matrix; we
  do not consider off-diagonal elements, which can be relevant for example to investigate  decoherence processes and quantum-to-classical transition. See e.g.  \cite{Brandenberger:1990bx,Albrecht:1992kf,Polarski:1995jg,Calzetta:1995ys,Lesgourgues:1996jc,Kiefer:1998qe,Burgess:2006jn,Kiefer:2008ku}.}:
\be
P_k^{(\lambda)}\,=\,|\Psi_k^{(\lambda)}|^2\,.
\ee
Using the relation \eqref{quadANps},
the  normalized probability reads
\be
P_k^{(\lambda)}\,=\,\frac{f_k}{\pi}
\,\exp\left\{
-f_k \left(  h^{(\lambda)}_k-g_0^\star\,\frac{\delta_{k0}}{f_k} \right)
\left(  h^{(\lambda)}_{-k}-g_0\,\frac{\delta_{k0}}{f_k} \right)
\right\}\,,
\ee
with
\bea
f_k&=&a^2(\tau)\left( \alpha_k(\tau)+\alpha_{-k}(\tau)\right)\,,
\\
g_0&=&a^2(\tau)\,\beta_0(\tau)\,=\,{\cal C}_\beta\,a^2(\tau)\,\alpha_0(\tau)
\,.
\eea

The probability density $P_k^{(\lambda)}$ is an important building block for the arguments we  develop next. In fact, we will work only in terms of probabilities for determining our stochastic
evolution equation.  By
differentiating along time, and by making use of the evolution equation \eqref{eqak}, it is straightforward to prove
that it satisfies a Fokker-Planck-like equation:
\be\label{eveqmk1}
\frac{\partial P^{(\lambda)}_k}{
\partial \tau}\,=\,\omega_{k}\,\frac{\partial^2\,P^{(\lambda)}_k}{\partial  h^{(\lambda)}_k\,\partial h^{(\lambda)}_{-k}}+\omega_{0}\,\,
 \left[ \frac{\partial}{\partial  h^{(\lambda)}_k} \left(
h^{(\lambda)}_k\,P^{(\lambda)}_k \right)
+
 \frac{\partial}{\partial  h^{(\lambda)}_{-k}} \left(
h^{(\lambda)}_{-k}\,P^{(\lambda)}_{-k} \right) \right]
\,,
\ee
with
\bea
\omega_{k}&=&\frac{i}{a^2(\tau)} \frac{\alpha_k-\alpha_0-\alpha_k^\star+\alpha_0^\star}{\alpha_k+\alpha_k^\star}\,,
\\
\omega_{0}&=&-i (\alpha_0-\alpha_0^*)
\,.
\eea
Using equations \eqref{defak} we can also reexpress the previous formula as
\bea
\omega_{k}&=&-\frac{|\gamma_0(\tau)|^2}{a^2(\tau)}\,\partial_\tau\,\left(\frac{|\gamma_k(\tau)|^2}{|\gamma_0(\tau)|^2} \right)
\,,
\\
\omega_{0}&=&-\partial_\tau\,\ln\left(\frac{|\gamma_0(\tau)|^2}{a^2(\tau)} \right)
\,.
\eea
This is our starting point for developing a  coarse-graining procedure  to describe the dynamics of  superhorizon modes.

\subsection{ Coarse-graining  superhorizon tensor modes}
\label{sub_coarse}

We now apply   the previous formulas to the development of   a convenient coarse-grained
tensor field at superhorizon scales, and its  corresponding stochastic evolution  equation. 

We start defining long-wavelength (time-independent) superhorizon fields    as a sum over 
Fourier modes, with a cutoff controlled by the comoving horizon scale $k_h$: 
\bea\label{defcg1}
h_{ij}(\vec x)&=&\frac{2}{M_{\rm Pl}\,L^{3}}\sum_\lambda \sum_{\vec k,\,|k|<k_h}\,
\,h^{(\lambda)}_k\,{\bf e}_{ij}^{(\lambda)}(\hat k)\,e^{i  \vec k \vec x}\,,
\eea
where  the cutoff scale is    (we call ${\cal H}\,=\,a'/a^2$)
\be \label{defkh}
k_h\,\equiv\,\sigma\, a (\tau) {\cal H}(\tau)\,,
\ee
and $0\le \,\sigma\,\le 1$ is a constant that quantifies what  fraction of long wavelength modes we include
in the coarse-graining procedure. 
We express the sum in \eqref{defcg1} in terms of the time independent $c$-numbers $h^{(\lambda)}_k$ we used
in the previous subsection for expressing the waveform $\Psi$. 
The  coarse-grained quantity $h_{ij}$ in eq \eqref{defcg1}  is time-independent and we expect it
 to be  stochastically distributed at superhorizon scales. Being built in terms of the abstract  $c$-numbers $h^{(\lambda)}_k$, it does not spoil the isotropy of the underlying space-time geometry.


In fact, 
the  coarse-grained quantity $h_{ij}$ of eq \eqref{defcg1} is a natural  definition of coarse-grained superhorizon tensor
mode, and we  use it in what follows. 
Being  constituted  by a combination of $h^{(\lambda)}_k$ modes at large scales $k\,<\,k_h$, we define the probability density related with the coarse-grained quantity $h_{ij}$ as
the product of the independent probabilities associated with each of the Fourier modes entering in eq \eqref{defcg1}:
\be\label{defCGP}
P(\tau,\,h_{ij}(\vec x))\,\equiv\,\, \Pi_\lambda\,\Pi_{|k|<k_h}\,P_k^{(\lambda)}
\,.
\ee
Notice that the product depends only on the size of the momenta, and not on their directions. 
Starting from the $P_k^{(\lambda)}$ evolution equation \eqref{eveqmk1} for any given mode $k$, it is straightforward to obtain
an evolution equation for $P(\tau,\,h_{ij})$. Selecting any given $k$, we first   multiply both sides of \eqref{eveqmk1}  for all the remaining probability densities $\dots P_{k-2}^{(\lambda)} \,P_{k-1}^{(\lambda)}\,P_{k+1}^{(\lambda)}\dots$.
Then, using eq \eqref{defCGP} and the Leibniz rule, we can reconstruct  an equation for $P(\tau,\,h_{ij}(\vec x))$.  

As shown in the technical appendix
\ref{appcogr},  the final result is a Fokker-Planck equation controlling the probability density $ P(\tau,h_{ij})$
\be
\label{finFP1}
\frac{1}{ a(\tau) \cal H(\tau)}\,
\frac{\partial P(\tau,h_{ij})}{\partial \tau}\,=\,{\cal N}(\tau) \,\frac{\partial^2 P(\tau,h_{ij})}{\partial h^2_{ij}}
+{\cal D}(\tau) \,\frac{\partial}{\partial h_{ij}} \left[ h_{ij}\,P(\tau,h_{ij}) \right]
\,.
\ee
The time derivative in the  left-hand side is assembled for convenience in the combination
\be
a\,{\cal H}\,d \tau\,=\,H\,d t\,=\,d n
\,,
\ee
with  $n$ the e-fold number
\be
\label{efNUM1}
n\,=\,\log{(a/a_*)}
\,,
\ee
a quantity that physically makes manifest the universe rate of expansion, and that
represents the physically correct time variable in the context of stochastic inflation \cite{Vennin:2015hra}.

The noise and drift in eq \eqref{finFP1} are given by \footnote{We pass to the continuous limit taking a large size $L$, and
expressing the sum as an integral: $(1/L^3)\,\sum_k\,=\,1/(2 \pi)^3\,\int d^3 k$.}
\bea
\label{exN1}
{\cal N}&=&\frac{2\,|\gamma_0(\tau)|^2}{M_{\rm Pl}^2\,\pi^2\,{\cal H}(\tau)\,a^3(\tau)}\,\int^{k_s}_{k_h} k^2 \,d k\,\partial_\tau\,\left(\frac{|\gamma_k(\tau)|^2}{|\gamma_0(\tau)|^2} \right)
\,,
\\
\label{exD1}
{\cal D}&=&-\frac{2}{{\cal H}(\tau)\, a(\tau)}\,
\partial_\tau\,\ln{\left(\frac{|\gamma_0(\tau)|^2}{a^2(\tau)} \right)}
\,.
\eea
%
Notice that while the drift depends on the zero mode only, the noise involves an integration 
over all the super-horizon modes, with an horizon-size lower cutoff given by $k_h\,=\,\sigma\,a {\cal H}$
as in eq \eqref{defkh}, and an upper  cutoff $k_s$ which controls the total size of the superhorizon region
experienced by the long modes. As we will see, the final results do not depend on $k_s$, at least
for physically relevant scenarios, hence  there are  no large infrared effects depending on the total size
of the superhorizon region. In fact, we can continuously reduce the size of $k_s$ in the final results, making it as small as we please. 
 
Up to an irrelevant constant overall factor, we can write 
\bea
|\gamma_0(\tau)|^2
&\propto&
a^2(\tau)\,\left\{ 1+ \sigma\,{\cal I}(\tau) \,\left[ 2  \,\cos{\Delta \theta}  + \sigma \,{\cal I}(\tau) \right] \right\}
\,,
\\
&=&a^2(\tau)\,\left\{ 1+ \sigma\,\Pi(\tau) \right\}
\,,
\eea
where ${\cal I}(\tau)\,=\,\int_{\tau_\star}^{\tau}\,
 {d \tilde \tau}/{a^2(\tilde \tau)}$ as given in eq \eqref{intItau}.  We
 introduce
\be
\Pi(\tau)\,\equiv\,
2 \,{\cal I}(\tau)   \,\cos{\Delta \theta}  + \mu \,{\cal I}^2(\tau) 
\,,
\ee 
with $\mu$ the  constant parameter appearing in the solution for the zero mode, see eq \eqref{sola0}. 
Substituting these expressions in formulas \eqref{exN1} and \eqref{exD1}  
we get
\bea
\label{genN1}
{\cal N}&=&\frac{2\,\left( 1+\mu\, \Pi(\tau)\right)}{M_{\rm Pl}^2\,\pi^2\,{\cal H}(\tau)\,a(\tau)}\,\int_{k_h}^{k_s} k^2 \,d k\,\,\partial_\tau\left(\frac{|\gamma_k(\tau)|^2}{
a^2(\tau)\,\left( 1+\mu\, \Pi (\tau)\right)
} \right)\,,
\\
\label{genD1}
{\cal D}&=&\frac{2}{{\cal H}(\tau)\,a(\tau)}\,
\partial_\tau\ln\left(\frac{1}{ 1+\mu\,\Pi(\tau)}\right)
\,.
\eea
It is also interesting to explicitly  consider  cases where  the decaying mode contribution
is set to zero, by selecting $\mu\,=\,0$. Then the drift vanishes, ${\cal D}=0$, since eq. \eqref{genD1}
is proportional to $\mu$. 
The noise instead simplifies to
\begin{equation}
\boxed{
{\cal N}\,=\,\frac{2}{\pi^2\,M_{\rm Pl}^2\,{\cal H}(\tau)\,a(\tau)}\,\int_{k_h}^{k_s} k^2 \,d k\,\partial_\tau\,\left(\frac{|\gamma_k(\tau)|^2}{
a^2(\tau)
} \right)}
\label{noiSIM1}
\end{equation}
a formula that plays an important role for our applications. 
\smallskip
\noindent
The following  physically relevant properties are worth emphasizing:
\begin{itemize}
\item
 Expressions \eqref{finFP1}, \eqref{genN1} and \eqref{genD1} are  general and valid for any cosmological space-time
 $a(\tau)$. Once we 
 have control on the expressions for $\gamma_k(\tau)$ for each $k$,
 we can compute -- analytically or numerically -- the  expressions for noise and drift \eqref{genN1}, \eqref{genD1}
 in a broad variety of physically interesting  situations.
 \item
  The noise ${\cal N}$ in eq \eqref{genN1} 
 is controlled  by a sum of time derivatives  of superhorizon modes, and
  depends on the time dependence of
   {\it all} the superhorizon modes $k_s\le k\le k_h$. 

 
 As we will see in the next section, such time dependence
 is a feature of  rapidly expanding space-times, and 
  the integral   \eqref{genN1} is associated with the 
 rate of change of the comoving horizon. This phenomenon controls the flow of modes crossing
 the horizon, and fits well with  the heuristic picture that a source for the stochastic noise ${\cal N}$  is
 due to       modes continuously
 crossing the cosmological horizon separating large and small scales. Effectively, we are
 dealing with an open system \cite{Burgess:2014eoa},
 and the flow of modes produces an analog  of Brownian motion
 at cosmological scales. 
 
  Importantly, the order of integration   in 
 eq \eqref{genN1} is from $k_h$ to $k_s$, and physically assumes that the  noise is due to the flow of modes crossing the horizon from subhorizon to superhorizon scales. In a case where the situation is reversed, as what happens during standard cosmological epochs  
 after inflation ends, the order of integration should be reversed for obtaining a noise with positive sign (see examples in section  \ref{sec_appl}). 
 
 Also, eq   \eqref{genN1}  can include  {\it additional} sources of noise  in the superhorizon regime, due to correlations among positive and negative frequency modes 
 with the same momentum $k$. As an example, noise can be generated  by 
   particle production at superhorizon scales after
   the transition between distinct cosmological space-times, as what happens
  between inflation and radiation domination.  As far as we are
  aware, this is the first time these phenomena are explored in the context
  of a stochastic approach to tensor fluctuations. 
   We will discuss explicit examples of these possibilities  in
 %
%
 %
 %
  section \ref{sec_appl}.

\item 
The drift term \eqref{genD1}   depends  on the physics of the zero mode $\gamma_0$:  the decaying mode 
 appearing in $\gamma_0$  introduces a dependence on initial conditions at early times,
 %
  through the coefficients of the integral ${\cal I}(\tau)$. 
 On the other hand, if  the cosmological evolution corresponds to an attractor, 
  ${\cal I}$ and $\Pi$ become rapidly a constant: all the effects of the decaying mode drop out from 
  expressions \eqref{genN1} and \eqref{genD1}, and the dynamics is  well described by Markovian
  evolution, independent from initial conditions. In this limit (or alternatively switching off the effects of the decaying mode  by selecting $\mu=0$) the drift vanishes, and the noise reduces to
  eq \eqref{noiSIM1}.
\end{itemize}

\subsection{  The   spectrum and entropy of
inflationary
 tensor modes}
\label{sub_spectrum}

\subsubsection*{The spectrum}

In many cosmological situations it is important to compute  the spectrum of tensor modes at horizon crossing. It
is straightforward to obtain its expression starting from the Fokker-Planck equation we derived. 
For simplicity we consider an attractor cosmological evolution, where the decaying mode becomes
rapidly negligible, and the drift vanishes. The Fokker-Planck equation \eqref{finFP1} corresponds to Einstein formulation 
of the theory of Brownian motion, and
reads (we express it in terms of the e-fold number, $dn \,=\, a {\cal H}\,d \tau$)
\be
\frac{\partial\,P}{\partial \,n}\,=\,{\cal N}(n)\,\frac{\partial^2\,P}{\partial h_{ij}^2}\,.
\ee
This equation
 can be easily integrated providing the Gaussian probability density, when assuming a positive ${\cal N}$
\be\label{solPR1}
P(n, h_{ij})\,=\,\frac{1}{\sqrt{2 \pi\,B(n)}}\,e^{-\frac{ h_{ij}^2}{2 B(n)}}\hskip1cm,\hskip1cm {\rm with}\,\,B'(n)\,=\,2 {\cal N}(n)\,.
\ee
Such probability density leads to the
two-point  function  for  superhorizon tensor modes as
\be  \label{def2pt1}
\langle  h_{ij}^2(\tau, x)\rangle\,=\,\int d  h_{ij}\, h_{ij}^2\,P\,=\,B(n)
\,.
\ee
The  two-point function 
depends on the e-fold number, and it is independent from the spatial position. The same quantity  $\langle  h_{ij}^2\rangle$ can
also be expressed in Fourier space, as an integral over the long wavelength tensor spectrum  up to the cutoff scale:
\be \label{def2pt2}
\langle  h_{ij}^2(\tau, x)\rangle\,=\,
\int_{{\ln k_{s}}}^{\ln k_{h}}{\cal P}_T\,d \ln k
\,.
\ee
To compute the value of the tensor spectrum at horizon scales
 we can use   equations \eqref{def2pt1} and \eqref{def2pt2} together, as discussed in \cite{Kunze:2006tu}, 
and use $d \ln k_{h}\,=\,d \ln (a {\cal H})$. 

We obtain
\bea
{\cal P}_T&=&\frac{d\,\langle h_{ij}^2(\tau, x)\rangle}{d\,\ln k_{h}}
\,,
\nonumber
\\
&=&\left( \frac{d n }{d  \ln (a {\cal H})} \right)\,\frac{d\,\langle h_{ij}^2(\tau, x)\rangle}{d\,n}
\,=\,\left( \frac{d n }{d  \ln (a {\cal H})} \right)\,\frac{d\,B(n)}{d\,n}
\nonumber
\,,
\\
&=&
 \frac{2\,a^2 {\cal H}^2}{\left| a^2 {\cal H}^2+a\,{\cal H}'\right|}\,
\,{\cal N}\,.
\label{posge1}
\eea
Hence~\footnote{In   cosmological
 phases following the end of inflation
the denominator of the overall coefficient in eq  \eqref{posge1}  would be negative, 
 in absence
of the absolute value. However, in these cases the flow of modes is from superhorizon to subhorizon scales (see comment in the second bullet point after eq \eqref{noiSIM1}). This fact changes our arguments here by an overall sign, leading to expression in eq \eqref{posge1} (with the absolute value).},
 knowing the profile of ${\cal N}$ as a function of the e-fold
number $n$, equation \eqref{posge1} provides the tensor spectrum at
horizon crossing. For the three cases of pure de Sitter, radiation domination, and matter
domination that we study next we find
\bea\label{posge2}
 \frac{2\,a^2 {\cal H}^2}{\left| a^2 {\cal H}^2+a\,{\cal H}'\right|}\,=\,\begin{cases} 
2 &\text{for de Sitter}\,,
\\
2
&\text{for radiation domination}\,,\\
4
&\text{for matter domination}\,.
 \end{cases}
\eea
It is interesting to compare it with the tensor spectrum deep at superhorizon
scales, computed with standard QFT methods (see e.g. \cite{Maggiore:2018sht}). One gets
\be
\label{ptQFT}
{\cal P}_T\,=\,\lim_{k\to0}\,\frac{4\,k^3}{\pi^2}\,\frac{|\gamma_k|^2}{a^2}\,,
\ee
where the modes $\gamma_k$ are solutions of eq \eqref{eqphk1}.
Notice that while eq \eqref{ptQFT} depends only on very large-scale modes with  $k\to0$, the
 stochastic prediction \eqref{posge1} depends on the noise ${\cal N}$ which
 involves a combination over all the superhorizon modes.  

\subsubsection*{The Gibbs entropy}


It is also interesting to compute the classical Gibbs entropy associated with our coarse-grained definition of superhorizon tensor fluctuations, see eq \eqref{defcg1}.
(See also \cite{Kiefer:1999sj} for a discussion  on the entropy of tensor fluctuations from inflation.)
From the expression \eqref{solPR1} (we set to one the Boltzmann constant), we get
\bea
S(n)&=&-\int d h_{ij}\,P(n,\,h_{ij})\,\ln \left[ P(n,\,h_{ij})\right]\,,
\\
\label{resGBen1}
&=& \frac12\,\ln\left[ B(n) \right]+{\rm constant}\,.
\eea
We learn that the entropy increases with the universe expansion, as long as the noise is positive:
\be
\frac{d S}{d n}\,=\,\frac{B'(n)}{2\,B(n)}\,=\,\frac12\,\frac{{\cal N}(n)}{\int^n  {\cal N}(n')\,d n'}
\,.
\ee
If the noise ${\cal N}$ is constant (or if it rapidly approaches a constant), then $B \propto 2 n$, and we find that  the rate of variation of the entropy is inversely proportional to the e-fold number: $d S/d n\,=\,1/(2 n)$. In our set up the Gibbs entropy grows logarithmically with the number of e-folds $n$: $S\propto (\log n)/2$. 
 
 Let us briefly discuss the conceptually important case of de Sitter space, and compare
  our coarse grained entropy with the Gibbons-Hawking entropy $S_{dS}\,=\,\pi\,M_{\rm Pl}^2/H_0^2$ (with $H_0$ the constant Hubble parameter).  
   As we are going to learn
 in section \ref{sec_appl}, the noise is constant in de Sitter, hence the coarse grained entropy  associated with superhorizon tensor modes  
 grows logarithmically  as $S_{cg}\,=\, \log{\sqrt{n/n_\star}}$, with $n_\star$
 a reference e-fold number.  $S_{cg}$ contributes to the energy budget, and keeps smaller than  $S_{dS}$ as long as $n\,\le\, e^{2S_{dS} }$, a  limit on the e-fold number imposed
 by the entropy bound. We point out however that here we only considered the coarse-grained Gibbs entropy, while we do not include entanglement effects that have
 been argued to contribute to the entropy budget by a function linearly growing with the number of e-folds: see \cite{Kiefer:1998qe}.

 \section{  Applications} 
 \label{sec_appl}

 The general formulas
 we obtained in the previous section 
  will now  be  applied to physically interesting cases, also
 in  contexts that are seldom considered in stochastic approaches to cosmological inflation. 
 
 In section \ref{sec_puredS} we use our stochastic approach to
 reproduce in this context well-known  QFT results for the spectrum of superhorizon tensor
 modes during inflation. We also go beyond the standard case, including in our  stochastic approach
a scenario with a phase of non-attractor evolution.

 In section \ref{sec_power} we consider the case of power-law inflation controlled
 by a parameter $\epsilon$ controlling  the departure from a pure
 de Sitter expansion. We show  that
 our formalism is sufficiently flexible to provide an exact, analytic
   expression for the noise  that reduces to the de Sitter one in
  the limit $\epsilon\to0$. We also show that the tilt $n_T$ of the tensor
  spectrum obtained by our stochastic method satisfies the expected
  consistency relation $n_T\,=\,-2\epsilon$. We derive an expression
  for the tensor spectrum using the stochastic formulation that does not require
  $\epsilon$ to be small.

 In section \ref{sec_radiation} we then consider   cosmological
 scenarios where  epochs of radiation and matter domination follow
 the phase of inflation. In this situation, we are interested
 to derive a stochastic formulation able to describe   superhorizon tensor  modes
 in the process of reentering the horizon after inflation ends.  We  find  that
 this flow of  modes from
 large towards small scales
 can be influenced by those genuinely   superhorizon modes  created  by  space-time
 curvature  during radiation and matter  dominated eras. We derive the corresponding expressions
 for the tensor spectrum at horizon exit, and 
compute the associated coarse-grained Gibbs entropy.  
 
  


\subsection{ Pure de Sitter expansion (plus an  extension to non-attractor evolution)}
\label{sec_puredS}

We start discussing the simplest  case of a  de Sitter universe, described by the conformal scale factor
\be
a\,=\,-\frac{1}{H_0 \tau}\,,
\ee
with $\tau<0$, and $H_0$ a constant of dimensions of inverse time, corresponding to the Hubble parameter
${ H}(\tau)\,=\,{a'}/{a^2}$. 
 For this choice of the scale factor, one has 
\be \label{secda1}
\frac{a''}{a}\,=\,\frac{2}{\tau^2}\,.
\ee
The mode function solving eq \eqref{eqphk1}, satisfying the Bunch-Davies conditions,  results

\be
\gamma_k\,=\,\frac{1}{\sqrt{2  k }}\, {e^{-i   k \tau}} \,
\times \left(1-\frac{i}{k \tau} 
\right)\,,
\ee
up to an overall phase that does not enter into the final results. 
 The integral ${\cal I}(\tau)$ of eq \eqref{intItau} controlling the effect of the decaying mode is  ($\tau_\star\le\tau\le 0$)
 \bea
{\cal I}(\tau)&=&\int_{\tau_\star}^\tau\,d \tau'\,H_0^2\,\tau'^2\,=\,\frac{H_0^2}{2}\,\left( \tau^3-\tau_\star^3 \right)\,,
\\
&=&-\frac{H_0^2\tau_\star^3}{2}\left(1-e^{-3 n} \right)\,,
\label{resDS}
\eea
where the number $n$ of e-folds is defined as
$
n\,=\,\ln \left[
{a(\tau)}/{a(\tau_\star)}
\right]
$. 
This implies that the quantity  ${\cal I}$ rapidly approaches a constant during inflation, and its contributions
to noise and  drift are exponentially suppressed: we can safely assume that contributions proportional
to $\sigma$ vanish in all our expressions. Hence  the drift contribution to the Fokker-Planck equation is zero in this limit. 
 For computing the noise we need the combination
 \bea
 \frac{|\gamma_k|^2}{a^2}&=&\frac{H_0^2}{2\,k^3}\,\left(1+k^2 \tau^2 \right)\,=\,
\frac{H_0^2}{2\,k^3}\,\left(1+\frac{k^2}{H_0^2\,a^2} \right)
\,,
 \nonumber
  \\
  \label{sqdS1}
 &=&\frac{H_0^2}{2\,k^3}\,\left(1+\frac{k^2}{H_0^2\,a_\star^2}\,e^{-2 \,n} \right)\,.
 \eea
Its time (or e-fold) dependence -- which controls the noise, see eq \eqref{noiSIM1} -- is limited to the second term inside the parentheses, characterizing  the rate of
change of the comoving horizon.

 Substituting eq \eqref{sqdS1} in the expression \eqref{noiSIM1} for the noise, we can easily perform the integral.  We get the expression
\be
\label{nopds1}
{\cal N}\,=\,\frac{H_0^2}{M_{\rm Pl}^2 \pi^2}\left(\sigma^2-k_s^2\,\tau_\star^2\,e^{-2\,n} \right)\,.
\ee
The result depends on the choice of the  cutoff $k_h\,=\,\sigma\,a {\cal H}$, and the infrared cutoff
$k_s$. For the  cutoff $k_h$ we choose $\sigma=1$: we include {\it all} the super-horizon modes starting from
horizon crossing, assuming all of them contribute in forming the noise. The choice of the infrared cutoff
$k_s$ is instead not important, since its contribution is exponentially suppressed as the e-fold number increases. After a few e-folds we then get the following expression:
\be
\label{nopds}
{\cal N}\,=\,\frac{H_0^2}{M_{\rm Pl}^2 \pi^2}
\,.
\ee
This is the expected 
 result for the noise coefficient. Indeed, using the fact that  $d n\,=\,d \ln{a {\cal H}}$ for a pure de Sitter evolution, eq \eqref{posge1} provides
 \be
 {\cal P}_T\,=\,\frac{2\,H_0^2}{ \pi^2\,M_{\rm Pl}^2}\,,
 \ee
 which is the well-known  spectrum of tensor modes at very large scales in the limit of pure de Sitter expansion,
   obtained using QFT methods and formula \eqref{ptQFT}. The statistics of the stochastic spectrum of coarse-grained modes maintains
 its properties from  horizon exit     up to very large scales, as expected given
 that the influence of decaying modes is negligible. 

\smallskip

We now briefly discuss how these classic results can change,  modifying one of the assumptions  made so far for  the case of pure de Sitter expansion.  
%
 %
During inflation, transitory phases of non-attractor can enhance 
the spectrum of fluctuations -- this mechanism
is particularly interesting in view of producing primordial
black holes (see e.g. \cite{Carr:2016drx,Sasaki:2018dmp} for reviews). 
 While this possibility has been mostly explored
 in the scalar sector, it might occur in the tensor
 sector as well \cite{Mylova:2018yap,Ozsoy:2019slf}. During non-attractor, the would-be decaying mode 
 proportional to the quantity $\int^\tau\,d \tilde \tau/a^2(\tilde \tau)$
 does not decay but grows.  Possible effects of this phenomenon
 for what respects quantum contributions to stochastic
 quantities have been explored in recent literature, see e.g. \cite{Biagetti:2018pjj,Ezquiaga:2018gbw,Cruces:2018cvq,Ezquiaga:2019ftu,Figueroa:2020jkf}. 
 
 Here we focus  our analysis  on understanding how a non-attractor regime
 influences the classical drift in the stochastic  Fokker-Planck equation,
 using the formalism we developed.

  The simplest possibility to consider is a model of non-attractor
 corresponding to a contracting universe, with $a(\tau)\,=\, a_0\,\tau^2$ ($a_0$
  is a normalization factor, and $-\infty<\tau<0$) 
 so
 that $a''/a\,=\,2/\tau^2$, as for the case of de Sitter (see eq \eqref{secda1}). This
 implies that the 
 solution for the mode function $\gamma_k$ is the same as in de Sitter
expansion. The number $n$ of e-folds of contraction is connected to the time variable
by  $\tau\,=\,\tau_*\,e^{-2 n}$, for $\tau_*<\tau<0$.

We find that the integral ${\cal I}$
 is
 \be
 {\cal I}\,=\,-\frac{1}{3 a_0^2\,\tau_\star^3}\left(e^{3 n/2}-1\right)\,,
 \ee
 so it exponentially  grows with the number $n$ of e-folds of contraction (instead
 of approaching a constant as in de Sitter, see eq \eqref{resDS}).   
  Calculating the drift as in eq \eqref{genD1},  in the limit of large e-fold number
  we find the expression
    \be
  {\cal D}\,=\,-6-6 \,e^{-3 n/2}\,\left(1+\frac{3\,a_0^2 \,\tau_\star^3\,\cos\Delta}{\mu}
  \right)+{\cal O}(e^{-3 n})\,.
  \ee
  So the drift
   approaches an order-one constant as contraction proceeds, and can
   influence considerably the stochastic evolution. It would be interesting to study
   more generally  stochastic features  of non-attractor inflation using our method: we
    postpone this investigation to future analysis.

\subsection{  Power-law expansion  }
\label{sec_power}


We now apply our formalism to power-law expansion, 
described by the scale factor
\bea
a(\tau)&=&-\frac{1}{H_0\,\tau^{1/(1-\epsilon)}}\,,
\eea
for constant $\epsilon$,
with de Sitter space corresponding to $\epsilon=0$. The parameter $\epsilon$ is associated with derivatives of the Hubble parameter ${\cal H}\,=\,(d a)/(a^2\,d \tau)$ through the  definition
\be
\epsilon\,=\,-\frac{1}{{\cal H}^2}\,\frac{d {\cal H}}{a\,d \tau}\,.
\ee

\begin{figure}[h!!]
\centering
  \includegraphics[width = 0.5 \textwidth]{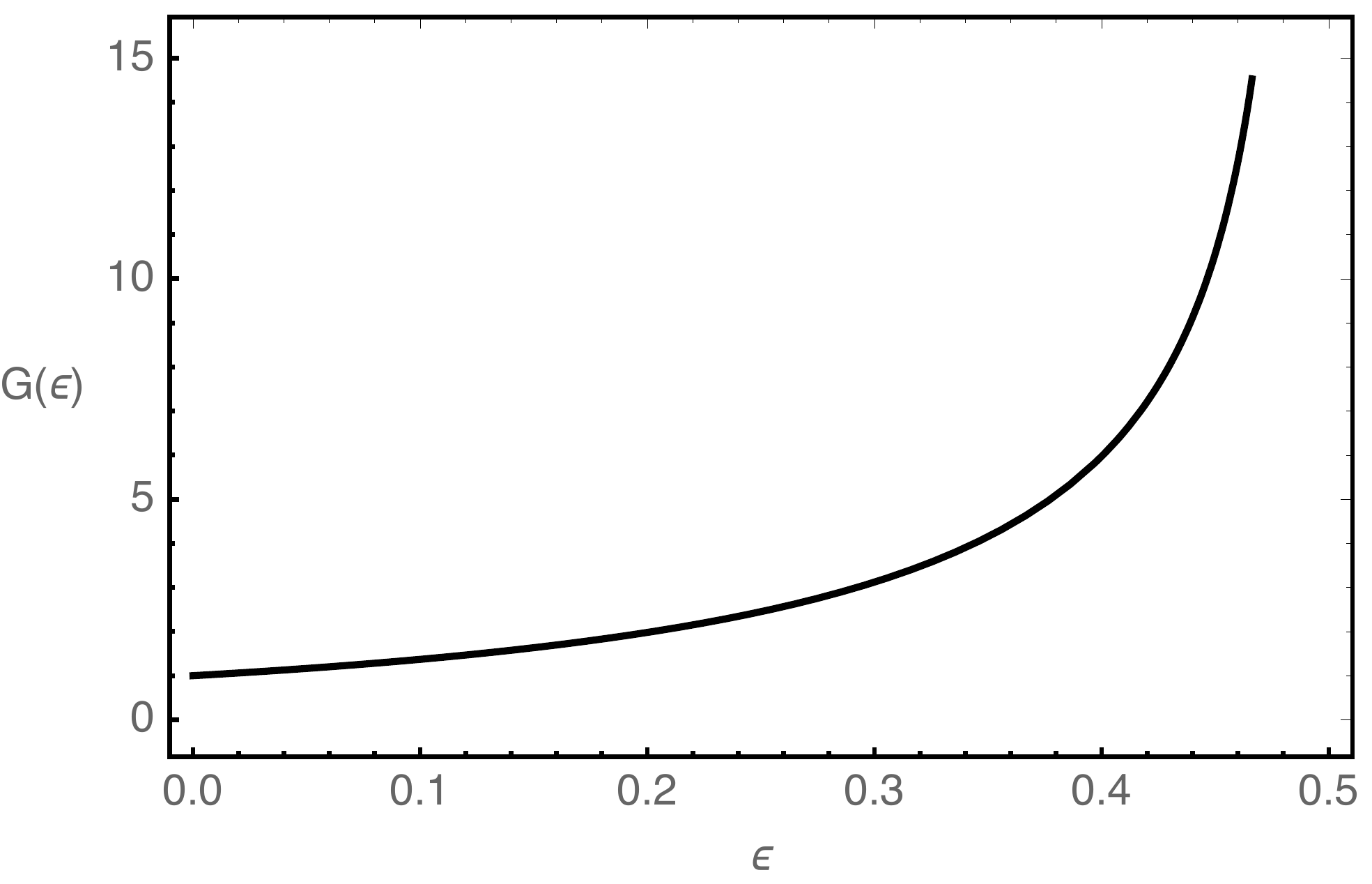}
 \caption{\it Plot of 
 the function $G(\epsilon)$ given in eq \eqref{defGe} as a 
  function of $\epsilon$.}
 \label{fig:plot2}
\end{figure}

  We are interested here
in cosmological space-times with $0\le \epsilon<1$. We can then express 
the Hubble
parameter and second time derivative of the scale factor as
\bea
{a { \cal H}}&=&-\frac{1}{(1-\epsilon)\,\tau}\,,
\\
\frac{a''}{a}&=&\frac{1-\epsilon/2}{(1-\epsilon)^2}\,\frac{2}{\tau^2}
\,.
\eea
%
It is a textbook exercise  to obtain the solution for the mode functions
that approaches a Bunch-Davies vacuum at early times -- see e.g. \cite{Maggiore:2018sht}. From such a solution one gets ($H_\nu^{(1)}(y)$ as the Hankel function
of the first kind)
\bea
\frac{|\gamma_k|^2}{a^2(\tau)}&=&\frac{\pi\,(-k \tau)}{2 k\,a^2(\tau)}\,| H_\nu^{(1)}( -k \tau) |^2\,,
\\
&=&\frac{\pi\,H_0^2}{2}\,\frac{(-k \tau)^{(3-\epsilon)/(1-\epsilon)}}{k^{(3-\epsilon)/(1-\epsilon)}}\,
| H_\nu^{(1)}( -k \tau) |^2\,,
\eea
where we denote
\be
\nu\,=\,\frac32\,\frac{1-\epsilon/3}{1-\epsilon}\,.
\ee
For calculating the noise, we assume that the UV cutoff is $k_h\,=\,a {\cal H}$, selecting $\sigma=1$ in eq \eqref{defkh}:
 in other words, as in section \ref{sec_puredS} we include all super-horizon modes in our definition of coarse-grained
 tensor quantity. The  value of $k_s$ is not important, since its contributions
 to the integral exponentially decay to zero as a function of  the e-fold number: in what follows for simplicity we set $k_s=0$.  
The noise is, always assuming  $\epsilon<1$,
\bea
{\cal N}&=&\frac{2}{M_{\rm Pl}^2\,\pi^2\,{\cal H}(\tau)\,a(\tau)}\,\int_{a {\cal H}}^0 k^2 \,d k\,\partial_\tau\,\left(\frac{|\gamma_k(\tau)|^2}{
a^2(\tau)
} \right)\,,
\\
&=&\frac{H_0^2}{ 2 \pi\,M_{\rm Pl}^2}\,(1-\epsilon)\,(-\tau)^{\frac{2 \epsilon}{1-\epsilon}}
\,\int_0^{1-\epsilon}\,{d (- k \tau)}\,{(-k \tau)^{\frac{-2 \epsilon}{1-\epsilon}}}
\,\frac{d }{d  (- k \tau)}\,\left( 
(- k \tau)^{\frac{3-\epsilon}{1-\epsilon}}
| H_\nu^{(1)}( - k \tau) |^2
\right)\,,
\nonumber
\\
&=&
\frac{2\,H_0^2\,{G}(\epsilon)}{ \pi^2\,M_{\rm Pl}^2}\,(-\tau)^{\frac{2 \epsilon}{1-\epsilon}}
\,,
\eea
where the overall coefficient ${G}(\epsilon)$ is given by
\bea\label{defGe}
G(\epsilon)&=&\pi (1-\epsilon)\,\int_0^{1/(1-\epsilon)}\,x^3\,d x\,\left[ J_{\frac{1+\epsilon}{2-\epsilon}} (x)
J_{\frac{3-\epsilon}{2-\epsilon}} (x)+Y_{\frac{1+\epsilon}{2-\epsilon}} (x)
Y_{\frac{3-\epsilon}{2-\epsilon}} (x)
\right]\,,
\eea
with $J_{\nu}(x)$, $Y_{\nu}(x)$ denoting respectively Bessel functions of the first and second kind. The function $G(\epsilon)$ tends to $1$ for $\epsilon$ small:
\be
G(\epsilon)\,\simeq\,1+2.94\,\epsilon+{\cal O}(\epsilon^2)\,,
\ee
and is represented in Fig \ref{fig:plot2}.

Since in this cosmological era the number of e-folds
is connected to time by ($\tau_*$ being a fiducial time $\tau_\star\le \tau\le 0$)
\be
\tau\,=\,\tau_\star\,e^{-n\,(1-\epsilon)}
\ee
We can then write the expression for the noise (choosing for definiteness $\tau_\star=-1$)
\bea
{\cal N}&=&\frac{H_0^2\,{G}(\epsilon)}{4 \pi^2}\,e^{-2 \epsilon\,n}\,
\eea
The corresponding tensor spectrum is given by formula
\eqref{posge1}. Its  tilt satisfies  the well-known relation
\be
n_T\,=\,\frac{d \,\ln{\cal P}_T}{d\,\ln k}\,=\,\frac{d \,\ln{\cal N}}{d\,n}\,=\,-2 \epsilon\,,
\ee
in agreement with standard QFT methods. 
\subsection{  From inflation to radiation and to matter  domination}
\label{sec_radiation}


After inflation ends, the  standard picture of  big bang cosmology  starts,  and 
the
universe enters in a phase of radiation followed by matter domination. Inflationary superhorizon modes 
reenter the horizon during these phases, and begin evolving and propagating through
cosmological distances. 
 During radiation or matter
domination, the stochastic distribution of superhorizon modes can  be described in terms of the physical arguments
we developed in the previous sections. The time-varying size of the cosmological horizon leads to a flow of modes back from superhorizon   to subhorizon scales -- 
  a process
 contributing to the stochastic noise  in the Fokker-Planck equation for our coarse-grained quantity. 
 In fact, we have an open system where
 the  Brownian motionlike phenomenon is  induced by the `holes' left
by the modes that leave the superhorizon regime.  Moreover, the transition from inflation to radiation domination
leads to  particle production at super-horizon scales, see e.g. \cite{Grishchuk:1974ny,Starobinsky:1979ty,Abbott:1985cu,Ford:1986sy}, and \cite{Maggiore:2018sht} for
a textbook discussion.  We might suspect that the superhorizon stochastic distribution  gets
affected by such phenomena. 
 
 \smallskip
 
In order to describe an universe where inflation (approximated
as de Sitter space) is followed by radiation domination, we parametrize the scale factor as 
\bea
a(\tau)&=& -\frac{1}{H_0 (\tau-\tau_0)} \hskip1cm \tau<0\,,
\\
a(\tau)&=& \frac{\tau+\tau_0}{H_0 \tau^2_0} \hskip1cm \tau>0\,,
\eea
for a continuous transition among the two regimes at $\tau=0$ ($\tau_0>0$ is a fiducial time). During radiation 
domination, the solution for mode function $\gamma_k$ is a linear combination of
plane waves
\be
\label{resmfrd}
\gamma_k\,=\,c_1(k) \,e^{i k \tau}+c_2(k) \,e^{-i k \tau}\hskip1cm \tau>0\,,
\ee
whose scale-dependent coefficients are determined by the
Israel conditions with inflationary modes in Bunch-Davies vacuum at $\tau<0$ 
\bea
c_1&=&\frac{e^{ i\,k \tau_0}\,
 }{\sqrt{8 \,k^5}\,\tau_0^2}
 \label{resc1}
 \,,
\\
c_2&=&-\frac{e^{ i\,k \tau_0}
}{\sqrt{8 
\,k^5}\,\tau_0^2} \left(1-2 i k \tau_0-2 k^2 \tau_0^2 \right)
 \label{resc2}
\,
\eea

We call negative frequency modes (in analogy with their
Minkowski counterparts)  the terms  weighted by $c_2(k)$ in eq  \eqref{resmfrd}. 
Their contribution 
 leads to particle production and amplification of of particle number at superhorizon scales.  We can compute the quantity entering in the noise
integrand in eq \eqref{noiSIM1}. We  get for $\tau>0$
\bea
\label{2gam2a}
\frac{|\gamma_k|^2}{a^2(\tau)}&=&\frac{H_0^2}{4\,k^5\,\left(\tau+\tau_0\right)^2}
\left[ 1+2\, k^4\,\tau_0^4+2 \,k \tau_0\,\sin{(2\, k\tau)} -\left(1-2 k^2 \tau^2 \right)\,\cos{(2\, k\tau)} \right]
\,.
\eea
 The oscillatory contributions within the  parentheses are due to interferences between positive
and negative frequency modes with the same $k$. Starting from expression \eqref{2gam2a},  using 
the definition in eq \eqref{ptQFT}, it is straightforward
to compute the spectrum of tensor fluctuations at late times $\tau/\tau_0\gg1$. We obtain \cite{Abbott:1985cu} \be\label{qftRD}
{\cal P}_T\,=\,\frac{2\,H_0^2}{\pi^2\,M_{\rm Pl}^2}\,\left(
\frac{\sin{k \tau}}{k \tau}
\right)^2\,,
\ee
at very large scales, $k \tau\ll1$, we find ${\cal P}_T\,=\,{2\,H_0^2}/{(\pi^2\,M_{\rm Pl}^2)}$. 

\medskip
We now analyze the problem from the perspective of the stochastic 
formalism developed in the previous sections. 
 The number of e-folds from the onset of radiation domination 
is
\be
\frac{\tau}{\tau_0}\,=\,e^n-1\,.
\ee
The
integral ${\cal I}$ of eq \eqref{intItau} results
\bea
{\cal I}(n)&=&\int_{\tau_0}^{\tau}\,\frac{d \tilde \tau}{a^2(\tilde \tau)}
\,,
\\
&=&\tau^3_0 H_0^2 \left(1-e^{-n} \right)
\,,
\eea
hence for increasing $n$  it
 approaches a constant, although more slowly than  in de Sitter space: the drift and
  the effects of the zero mode
 nevertheless are suppressed after a few e-folds, and we neglect them. 
Starting from eq \eqref{2gam2a}, it is straightforward  to perform  the analytic integrations~\footnote{Recall that we are in a situation where the flow of modes is from superhorizon to subhorizon 
scales, hence we should place an overall minus sign in eq  \eqref{noiSIM1}, as explained
in the second  point after that formula.}
associated with the noise of eq \eqref{noiSIM1}. 
A mixing between positive and negative frequency modes of  momentum $k$ is induced by the square 
of the mode function $|\gamma_k|^2/a^2(\tau)$, and leads to interesting effects.
%
%

The noise ${\cal N}$ controlling superhorizon modes during
 radiation domination
  is computed by the integral in eq \eqref{noiSIM1}, choosing $k_h\,=\,\sigma a {\cal H}$,  and leaving an arbitrary small $k_s$ as infrared cut-off. The quantity ${\cal N}(n)$ written
   as a function of the e-fold number  results
   
   \bea
{\cal N}(n)&=&\frac{H_0^2}{\pi^2\,M_{\rm Pl}^2}\,\frac{\sin^2 \sigma}{\sigma^2} 
\nonumber
\\
&&\times
\Big\{ 1
-\frac{H_0^2 \,\sigma^2\,e^{-2 n}}{2\,k_s^2\,\sin^2 \sigma} 
\big[ 1-2 k_s^4 \tau_0^4-\left(1-2 k_s^2 \tau_0^2 \right)
\cos{\left( 2 k_s \tau_0 \left( 1-\frac{e^n}{H_0 \tau_0}\right)\right)}
\nonumber\\
&& \hskip0.8cm
-2 k_s \tau_0 \,\sin{\left( 2 k_s \tau_0 \left( 1-\frac{e^n}{H_0 \tau_0}\right)\right)}
\big]
\Big\}+{\cal O}(e^{-3 n})
\,,
\label{expnoRD}
\eea
where the quantity ${\cal O}(e^{-3 n})$ decays at least as fast as $e^{-3 n}$ with the number of e-folds of evolution. We checked that
the complete expression for the noise is continuous when sending $k_s\to0$. Notice that the explicit dependence
on $k_s$ of eq \eqref{expnoRD} is exponentially suppressed with the e-fold number: after a few e-folds, the noise
approaches  a constant given by
   \bea
   \label{expnoRD2}
{\cal N}(n\gg1)&=&\frac{H_0^2}{\pi^2\,M_{\rm Pl}^2}\,\frac{\sin^2 \sigma}{\sigma^2}
\,.
\eea
Using eqs \eqref{posge1}, \eqref{posge2}, we  find for the tensor spectrum at horizon crossing, evaluated
after few e-folds of expansion, results 
\be\label{PTstochRD}
{\cal P}_T(n\gg1)\,=\,\frac{2\,H_0^2}{\pi^2\,M_{\rm Pl}^2}\,\frac{\sin^2 \sigma}{\sigma^2}
\,.
\ee
When  $\sigma \ll 1$,
the quantity in eq \eqref{PTstochRD} coincides with the QFT tensor spectrum of eq \eqref{qftRD} when evaluated deep at superhorizon scales $ k \tau\to0$. Physically, this choice for $\sigma$  implies that we  include in the coarse-graining procedure only modes
at very large scales, well beyond the horizon size -- see our definition of UV cut-off in eq \eqref{defkh}. This result  is intuitively 
  clear since the QFT approach focuses precisely on modes  at very large scales, hence
  there is no surprise that in this limit the two approaches agree.   
  Spanning the value of $\sigma$ within the interval $0\le \sigma\le1$, the size of the spectrum reduces of around 70\% with respect
to its $\sigma=0$ value. We interpret this suppression as due to  interference effects among modes spontaneously created  by space-time gradients at superhorizon scales, and    modes flowing from super to subhorizon scales. 
 Such interference is reduced when coarse graining only over modes  deep in the superhorizon regime, $\sigma\to0$. 


\begin{figure}
\centering
  \includegraphics[width = 0.45 \textwidth]{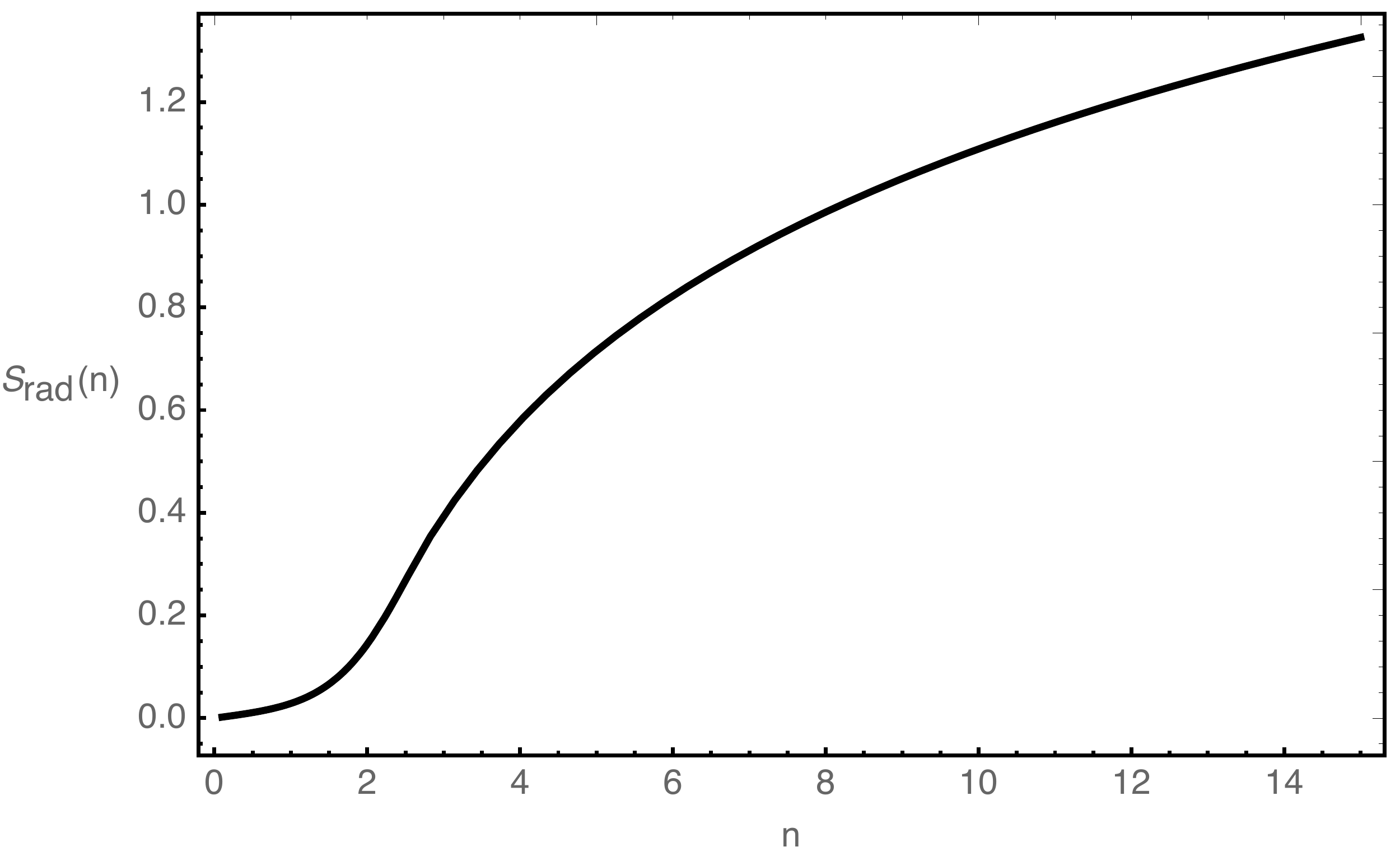}
    \includegraphics[width = 0.45 \textwidth]{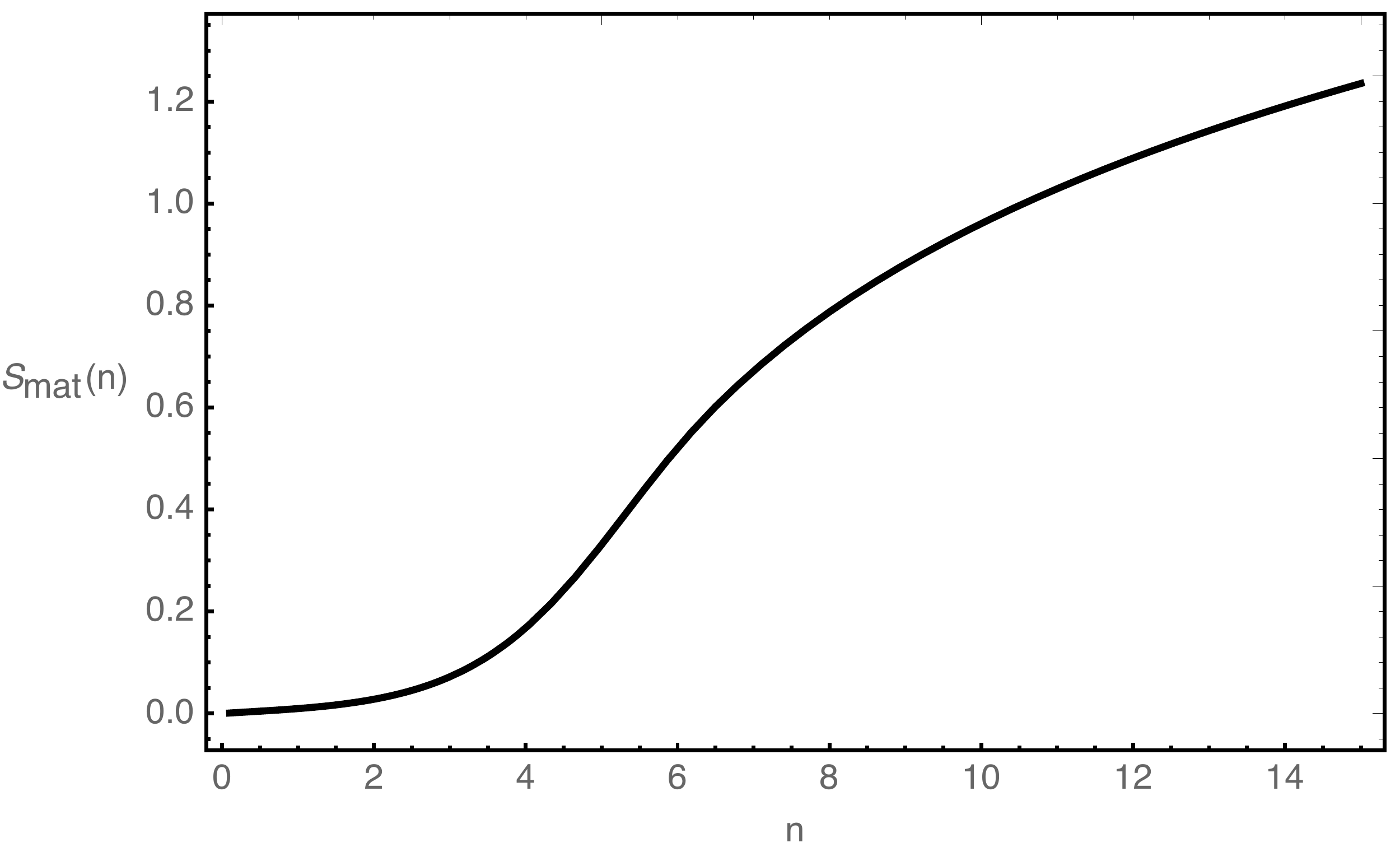}
 \caption{\it Plot of 
 the 
 coarse-grained entropy of  superhorizon
 tensor modes as a function of the e-fold number.
 We apply
 eq \eqref{resGBen1} to epochs  of radiation and matter
  domination. The integration constant  of eq  \eqref{resGBen1} 
 has been chosen in such a way  that the entropy vanishes at $n=0$. We choose $H_0\,\tau_0\,=\,4$, and $k_s \tau_0\,=\,0.1$, $\sigma=0.02$. In both cases the entropy scales as $\ln{(n^{1/2})}$ for large $n$.
 }
 \label{fig:plot1}
\end{figure}

\bigskip

We can also consider the case where a phase of matter domination follows the radiation-dominated era considered
above. The scale factors in the three epochs read ($\tau_0>0$ and $\tau_b>0$) 
\bea
a(\tau)\,=\,\begin{cases} 
-\frac{1}{H_0 (\tau-\tau_0)} &\text{for $\tau\le-\tau_b$}
\,,
\\
\frac{\tau+\tau_0+2 \tau_b}{H_0 (\tau_b+\tau_0)^2} &\text{for $-\tau_b\le \tau\le0$}
\,,
\\
\frac{(\tau+2\tau_0+4 \tau_b)^2}{4 H_0 (\tau_b+\tau_0)^2(\tau_0+2 \tau_b)} &\text{for $\tau \ge0$}
\,,
\end{cases}
\eea
and are continuous with their first derivative continuous at the transition epochs.
In the limit of very short radiation-dominated era, $\tau_b/\tau_0\ll1$, the solution for the mode function in matter domination for $\tau\,\ge\,0$ reads 
\be
\gamma_k\,=\,\frac{3}{8\sqrt{2}}\,\frac{e^{-i k (\tau+\tau_0)}}{\tau_0^3\,k^{9/2}}\,
\frac{(i+ k \tau_0) (-i+k \tau+2 k \tau_0)}{\tau+2 \tau_0}
\,.
\ee
Proceeding as above in the radiation-dominated case, we find the following expression for the noise as function 
of e-folds in the matter-dominated era, $n\ge0$
\bea
{\cal N}(n)&=&\frac{9 H_0^2}{128\,\pi^2\,M_{\rm Pl}^2} \frac{
\left(\sin{(2\sigma)}-2 \sigma\,\cos{(2 \sigma)}\right)^2}{\sigma^6}
\nonumber
\\
&&+\frac{9 H_0^2}{512\,\sigma^4\,\pi^2\,M_{\rm Pl}^2}\,e^{-n}\,\left[ 
3+16 \sigma^2+(8 \sigma^2-3) \cos{(4 \sigma)}
-12 \sigma \sin{(4 \sigma)}
\right]
\nonumber
\\
&&+{\cal O}(e^{-2 n})
\,.
\eea
After few e-folds of matted-dominated expansion, the noise approaches a constant. In this limit, using eqs
 \eqref{posge1} and \eqref{posge2}, we find the tensor spectrum at horizon crossing
 \be
 {\cal P}_T\,=\,\frac{18 H_0^2}{\pi^2\,M_{\rm Pl}^2} \frac{
\left(\sin{(2\sigma)}-2 \sigma\,\cos{(2 \sigma)}\right)^2}{(2 \sigma)^6}
\,,
 \ee
which approaches the standard large-scale value $ {\cal P}_T\,=\,2 H_0^2/(\pi^2 M_{\rm Pl}^2)$ in the limit $\sigma\ll1$. As for the
case of radiation  domination, the limit of small-$\sigma$ implies the inclusion only of  
 very large-scale modes  in the coarse-graining procedure. In spanning through the interval $0<\sigma<1$,  $ {\cal P}_T$ monotonically decreases,  reducing  to a size of 
 43 \% with respect to the $\sigma=0$ value. 
 
 \smallskip
 
 We conclude with few words about the behaviour of the coarse-grained Gibbs entropy, as derived in eq \eqref{resGBen1}. 
 Both for the cases of radiation and matter domination the noise approaches a constant
 as the cosmological evolution  proceeds, and the number of e-folds increases. In the transition
 between inflation and matter domination the noise has a richer profile: we use it for plotting
 the expression of the entropy in Fig \ref{fig:plot1}. We notice that in both cases the entropy increases as function of the e-fold number, with a steep slope for $n$ between  $2$ and $6$. Then, for large $n$,
 the entropy scales as   $\ln{(n^{1/2})}$, as expected. 

\section{Conclusions}
\label{sec_conc}

We discussed a coarse-grained prescription 
for describing the stochastic  superhorizon dynamics of inflationary
tensor modes, which seed 
the spectrum of primordial gravitational waves from inflation.  
We made precise the intuitive idea that the stochastic distribution of tensor
fields at superhorizon scales is due to the flow of  tensor modes between subhorizon and superhorizon scales.   
Our aim 
was to put in a firm footing a  consistent description of 
 inflationary  tensor modes   which allows one to deal with large infrared
 effects that characterize the dynamics of light fields in inflation.

Using basic principles
of quantum mechanics, we showed how the probability density
for the coarse-grained tensor modes satisfies a stochastic Fokker-Planck
equation, whose noise and drift are computable and depend
on the cosmological system under consideration.  The evolution
is well described by a standard Markovian process 
if the cosmological expansion follows an attractor,  and we also considered
how the dynamics is affected by the presence of non-attractor eras.
 Our stochastic formulas are applied to a variety
of cosmological frameworks, also cases not often  considered in the context
of stochastic inflation. We  obtained the expected results for noise
and drift in pure de Sitter and power-law inflation. But we also 
 explored  consequences of non-attractor phases as for example a contracting universe. Most
 notably,  we considered a cosmological space-time with transition from de Sitter (inflationary)
phase to radiation and to matter domination. This is the first time this topic is discussed  in the context of a stochastic approach
to superhorizon tensor modes.  
 The computation of the stochastic noise  made manifest   
  interference effects among 
  the flow of modes reentering the horizon after inflation ends,   and the superhorizon      
modes semiclassically  produced at large scales by  large space-time gradients.
 The formula for the noise depends on the number of e-folds of cosmic evolution, and it rapidly
 approaches   a constant value after few e-folds of expansion. 
   We  
 proved that our final
results do not depend on the choice of  infrared cutoff.
  Our stochastic results are then compared with the standard predictions of QFT applied
  to cosmology. The two  approaches give the same
  results for the power spectrum of tensor fluctuations if the coarse-graining procedure
  includes only modes deep in the superhorizon regime. This fact is intuitively 
  clear since the QFT approach focusses precisely on modes  at very large scales. We also quantitatively computed 
   the effect of including a larger portion of superhorizon modes in the coarse-graining
   prescription, showing that it 
  can change the amplitude of the tensor spectrum at horizon crossing  by an overall numerical coefficient   of order one. Hence, depending on the prescription adopted, the predictions
  for the amplitude of the spectrum can change. It would be interesting to further  consolidate our   
  physical understanding  of this fact, and its phenomenological consequences.




\smallskip
This work contains various novel results both for  developing a stochastic
approach to tensor fields from inflation, and for applying it to a variety of cosmological
settings.
 It would be interesting to further develop this approach 
 to better understand how much our quantitative results for noise and drift depend on the detailed
features  of cosmological space-times,  and on the transitions among different cosmological eras.
  It would also be
interesting to include self-interactions  (cubic or higher) among tensor fluctuations, and
then also include the effects of scalar perturbations in the analysis.

\subsection*{Acknowledgments}
It is a pleasure to thank Cliff Burgess and Ivonne Zavala for useful comments. GT is partially supported by the STFC grant ST/T000813/1.
\begin{appendix}

\section{ Fokker-Planck equation and coarse graining}\label{appcogr}

In this appendix we show  how to make use of our coarse-graining procedure
to pass from the evolution equation \eqref{eveqmk1} for a single-mode $k$ to the coarse-grained Fokker-Planck
equation eq \eqref{finFP1}.  The idea is to
  multiply both sides of \eqref{eveqmk1} -- defined for a certain fiducial  mode $k$ -- 
for all the remaining probability densities $\dots P_{k-2}^{(\lambda)} \,P_{k-1}^{(\lambda)}\,P_{k+1}^{(\lambda)}\dots$ of the remaining modes.
Then, using the definition of eq \eqref{defCGP}, we reconstruct  an evolution equation for $P(\tau,\,h_{ij}(\vec x))$.  To do so, we
we also need the fact that
\be
\frac{\partial h_{ij}(\vec x)}{\partial h^{(\lambda)}_k}\,=\,\frac{2
\,e^{i  \vec k \vec x}\,{\bf e}_{ij}^{(\lambda)}(\hat k) 
}{M_{\rm Pl}\,L^{3/2}}\hskip0.5cm;\hskip0.5cm
\frac{\partial h_{ij}(\vec x)}{\partial h^{(\lambda)}_{-k}}\,=\,\frac{2
\,e^{-i  \vec k \vec x}\,{\bf e}_{ij}^{(\lambda)}(\hat k) 
}{M_{\rm Pl}\,L^{3/2}}
\,.
\ee
We proceed to discuss this procedure analyzing its consequences for each term of eq  \eqref{eveqmk1}.
\begin{itemize}
\item[i)] The time derivative in the left-hand-side  (LHS). We multiply the  LHS  of eq  \eqref{eveqmk1} by all the the $P_{k'}^{(\lambda)}$ with
$k'\neq k$. We sum over momenta (positive and negative) and polarizations, obtaining
\be
 \sum_\lambda\,\sum_k\,\dots P_{k-1}^{(-\lambda)}\,P_{k-1}^{(\lambda)}\,P_k^{(-\lambda)}\,\frac{\partial P_k^{(\lambda)}}{\partial \tau}\,   P_{k+1}^{(\lambda)}\,P_{k+1}^{(-\lambda)}\dots\,=\,\frac{\partial P}{\partial \tau}\,  .
\ee
\item[ii)] The first derivatives in the right-hand-side  (RHS). We  multiply the  RHS  of eq  \eqref{eveqmk1} by all the  $P_{k'}^{(\lambda)}$  with
$k'\neq k$. We sum over momenta (positive and negative) and polarizations, obtaining 
\bea
&&
\sum_\lambda\,
\,\sum_k\,\dots P_{k-1}^{(-\lambda)}\,P_{k-1}^{(\lambda)}\,\omega_0\,\left[h_k^{(\lambda)}\,
\frac{\partial}{ \partial h_k^{(\lambda)}} \left(
P_k^{(\lambda)} \right)
 \right] \, P_k^{(-\lambda)} \, P_{k+1}^{(\lambda)}\,P_{k+1}^{(-\lambda)}\dots\,=\,
 \nonumber\\
 &=& \omega_0\,\sum_\lambda\,\sum_k\left[
h_k^{(\lambda)}\,\frac{\partial P}{ \partial h_k^{(\lambda)}} \right]
\,,
  \nonumber\\
 &=& \omega_0\,\sum_\lambda\,\sum_k\left[
 h_k^{(\lambda)}\,\frac{\partial h_{ij}}{ \partial h_k^{(\lambda)}}
 \right] \,
 \frac{\partial P}{ \partial  h_{ij}}\,,
   \nonumber\\
 &=&\omega_0\,h_{ij}\,\frac{\partial\,P}{ \partial  h_{ij}}\,,
\eea
since we recall $\omega_0$ is independent from $k$. In the previous expression,  we sum over indexes $ij$.
\item[iii)] The second derivatives in the RHS. We  proceed as before, and we express it as
\bea
&&
\sum_\lambda\,
\,\sum_k\,\dots P_{k-1}^{(-\lambda)}\,P_{k-1}^{(\lambda)}\,\omega_k\,\left[
\frac{\partial^2 P_k^{(\lambda)} }{ \partial h_k^{(\lambda)}  \partial h_{-k}^{(\lambda)}} 
 \right] \, P_k^{(-\lambda)} \, P_{k+1}^{(\lambda)}\,P_{k+1}^{(-\lambda)}\dots\,=\,
 \nonumber\\
 &=& \,\sum_\lambda\,\sum_k
 \omega_k\,
 \frac{\partial^2  P }{ \partial h_k^{(\lambda)}  \partial h_{-k}^{(\lambda)}} 
 \,,
  \nonumber\\
  &=& \,\sum_\lambda\,\sum_k
 \omega_k\,\frac{\partial h_{ij}}{\partial h_k^{(\lambda)}}\,\,\frac{\partial h_{ij}}{\partial h_{-k}^{(\lambda)}}
 \,\frac{\partial^2\,P}{\partial h_{ij}^2}
\,,
  \nonumber\\
    &=&\left(\sum_k \omega_k
\right)\,\left(\sum_\lambda {\bf e}^{(\lambda)}_{ij} {\bf e}^{(\lambda)}_{ij} \right)
\,\frac{\partial^2 P}{\partial h_{ij}^2}
\,,
   \nonumber\\
  &=&2\,\left(\sum_k \omega_k
\right)\,\frac{\partial^2 P}{\partial h_{ij}^2}
\,.
\eea
\end{itemize}
Collecting the  results we just obtained, we can write  a stochastic Fokker-Planck evolution equation for the coarse-grained
probability $P$, which reads
\be
\frac{1}{a \,{\cal H}}\,\frac{\partial P}{\partial \tau}\,=\,{\cal N}\,\frac{\partial^2\,P}{\partial h_{ij}^2}
+{\cal D}\,\frac{\partial}{\partial h_{ij}} \left( h_{ij}\,P \right)\,,
\ee
and noise and drift given in the main text: see eqs \eqref{genN1} and \eqref{genD1}. 
\end{appendix}


\providecommand{\href}[2]{#2}\begingroup\raggedright\endgroup

\end{document}